# Characteristics of a 9 MeV electron beam for total skin electron radiotherapy evaluated in comparison to a 6 MeV electron beam

T. Kairn[1,2,3][0000-0002-2136-6138], C. M. Lancaster[1][0000-0002-8797-9586] and S. B. Crowe[1,2,3][0000-0001-7028-6452]

[1] Cancer Care Services, Royal Brisbane and Women's Hospital, Herston, Qld, Australia
[2] School of Chemistry and Physics, Queensland University of Technology, Brisbane, Qld, Australia
[3] School of Information Technology and Electrical Engineering, University of Queensland, St Lucia, Qld, Australia
t.kairn@gmail.com

**Abstract.** Characteristics of a 9 MeV electron beam were investigated, to evaluate potential benefits for patients with extensive mycosis fungoides lesions that are deeper than commonly treated with 6 MeV total skin electron therapy (TSET or TSE/TSEI/TSEB/TSEBT). A comprehensive commissioning measurement program was completed for TSET delivery using high-dose-rate 6 MeV and 9 MeV electron beams from a Varian TrueBeam linac. Various phantoms and dosimeters were set up 300 cm from isocenter, behind a 6 mm PMMA spoiler screen, and used for optimising beam-pair gantry angles, measuring depth-dose, lateral and horizontal profiles and B-factors, as well as performing end-to-end tests. The key clinical distinctions observed for the 9 MeV TSET beam pair compared to the 6 MeV TSET beam pair were that the 80% dose depth increased to nearly 1 cm and the practical range increased to 3.5 cm, while the maximum dose remained within 0.2 cm of the surface. Vertical and horizontal dose profiles measured with the two energies were almost indistinguishable. The 9 MeV TSET beam has been shown to achieve greater depth penetration compared to the commonly used 6 MeV TSET beam, without detrimentally affecting the dose uniformity achievable in the patient plane. TSET treatments with 9 MeV electrons may be advisable for patients with extensive lesions that are too deep for effective treatment with a 6 MeV beam.

## 1 Introduction

Mycosis fungoides is a common form of cutaneous T-cell lymphoma that manifests as superficial plaques that can cover large areas of the patient's skin [1]. Mycosis fungoides can be effectively treated using total skin electron therapy (TSET, also known as TSE, TSEI, TSEB, TSEBT) which is commonly delivered today using a nominal 6 MeV electron beam, following a substantial literature recommending the use of electron beams in the energy range 4-8 MeV [1]. Spoiler screens and an extended source-to-surface distance (SSD) are also used to scatter and reduce the energy of the electron beams when they reach the patient [2, 3]. The use of these lower-energy electron beams has been preferred due to the commonly shallow depth of mycosis fungoides lesions, although energies up to 12 MeV may be used for boosts or node treatments [1].

Occasionally, however, patients present with extensive mycosis fungoides lesions that extend beyond a depth that can be effectively treated with the 6 MeV beam, so that the TSET treatment would not be justified by an appreciable clinical benefit. Historically, these rare cases may have been treated only with localized higher-energy electron treatments that cover only a limited number of smaller lesions, or turned away from radiation oncology departments entirely. For the Royal Brisbane and Women's Hospital, the recent installation of two Varian TrueBeam linacs (Varian Medical Systems, Palo Alto, USA) with the capability to deliver high-dose-rate electron beams at energies of both 6 MeV and 9 MeV, raises the possibility of implementing an additional TSET option, for patients with unusually deep extensive mycosis fungoides lesions.

This study therefore used a comprehensive set of commissioning tests to evaluate the characteristics of the 9 MeV high-dose-rate electron beam in comparison to the pre-existing 6 MeV high-dose-rate electron treatment option, for delivering TSET treatments to a broader cohort of mycosis fungoides patients.

## 2 Method

### 2.1 TSET treatment overview and measurement setup

A broadly consistent TSET treatment method has been in use in our department since the 1980s [4], using electron beam energies that have increased from 4 to 5 to 6 MeV, as new linac technology has become available. This method uses paired electron beams, delivered at a high dose rate, to treat patient who is standing behind a spoiler screen at extended SSD, using the six established "Stanford poses" to ensure maximum skin coverage [3]. The paired beams consistently use a field size of 36 ×36 $cm^2$ and a zero-degree collimator angle. Treatment MU calculations are performed manually (since patients cannot be CT scanned in their vertical treatment positions [5]) using data provided through the TSET commissioning process. These dose calculations are ultimately verified by performing in vivo dosimetry measurements at every fraction for every patient treated with TSET [6].

The data specifically required for treatment dose calculations include: a ratio that relates dose from the treatment field at extended SSD behind the spoiler screen to dose in a reference field measured using the IAEA TRS 398 protocol [7]; a correction for variations in patient position away from the extended-SSD reference plane; a prescription depth correction; and the B-factor, a ratio that relates the measurement in the extended-SSD reference setup to the mean dose around the patient's skin surface [3]. Relative dose profile measurements across the width and height of the patient treatment plane are also useful for providing guidance regarding expected TSET dose homogeneity. Finally, before all of the other measurements can be completed, the optimal gantry angles for delivering the paired electron fields using the 6 and 9 MeV high-dose-rate electron beams need to be identified.

While the electron reference dosimetry measurements are performed in a scanning water phantom with its surface at isocenter distance [7], all of the TSET-specific measurements listed above are performed in and around a consistent extended-SSD reference setup. Figure 1(a) illustrates the TSET reference setup, showing the relative locations of the linac isocenter (labelled as "Ref. pt on floor"), the 6 mm thick polymethyl methacrylate spoiler screen, and the wooden frame that both assists in TSET patient setup and is used to support measurement devices at the reference position (3000 mm lateral to isocenter at the same height as isocenter) and at other measurement positions needed for TSET commissioning. Figure 1(b) provides a potentially clearer illustration of how all these parts work together in practice.

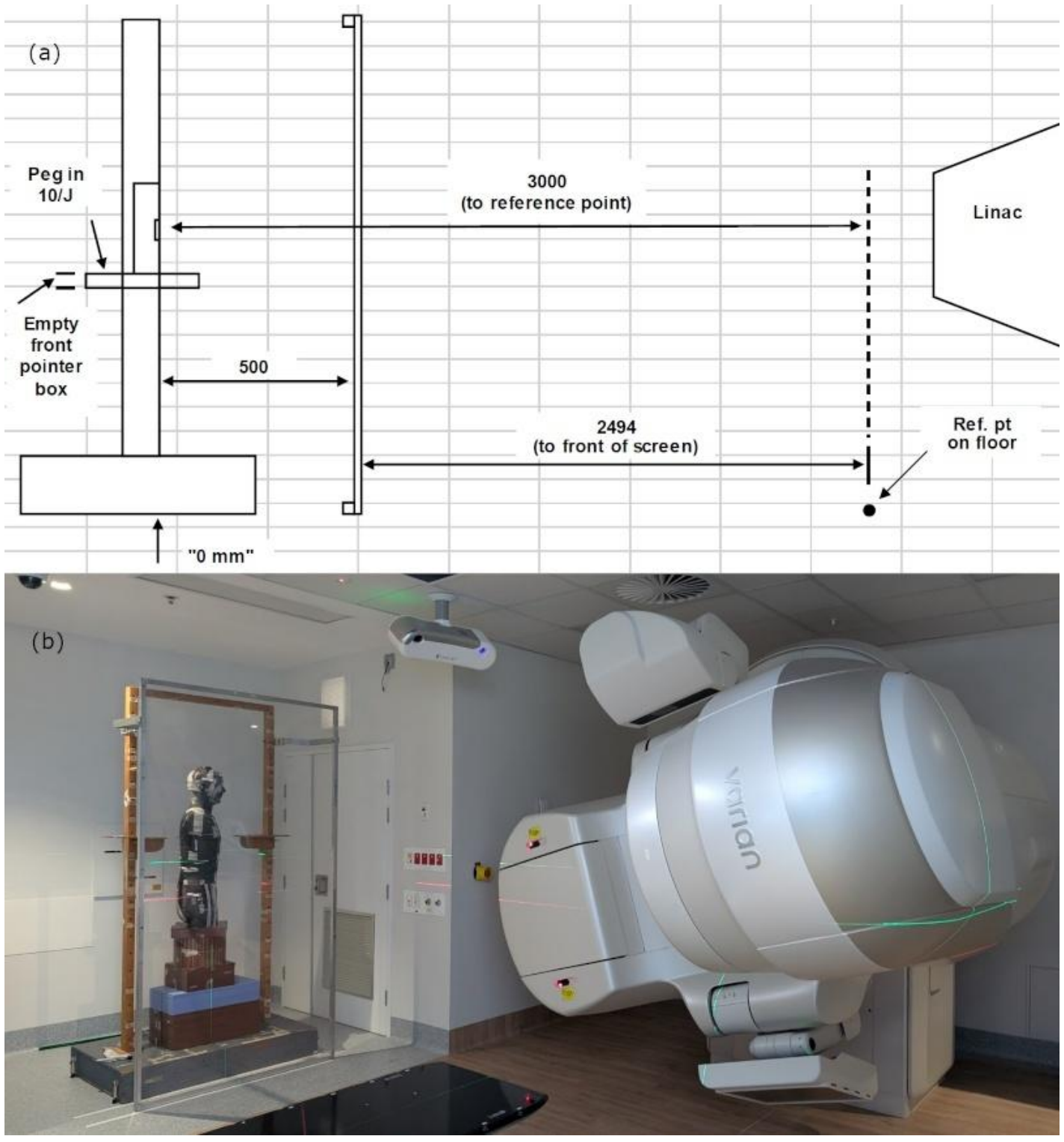

**Fig. 1.** (a) Diagram of extended-SSD reference setup for TSET dosimetry (all distances are in mm and "Ref. pt on floor" is directly below the linac isocenter, diagram is not to scale) and (b) photograph taken during TSET delivery to humanoid phantom.

## 2.2 Gantry angle selection

To identify the optimal paired gantry angles for delivering TSET treatments using the 6 and 9 MeV high-dose-rate electron beams, the measurement setup shown in Figure 1 was used, with a Roos chamber positioned at isocenter height, 3000 mm from the isocenter, in a 200 × 200 × 100 mm$^3$ Solid Water block (Gammasonics, now Sun Nuclear Corp, Melbourne, USA). Sheets of 200 × 200 mm$^2$ Goetingen RW3 White Water (PTW-Freiburg GmbH, Freiburg,

Germany) were placed over the Roos chamber, to provide sufficient electron build-up to achieve measurements at maximum dose depth.

Using 36 ×36 $cm^2$ fields with a zero-degree collimator angle, gantry angles between 15 and 20 degrees above and below horizontal were iteratively tested, seeking a pair of gantry angles that each produced readings that equated to 50% the readings from a corresponding horizontal beam.

### 2.3 Reference output measurements and B-factors

Extended-SSD reference output measurements were completed using the same measurement setup as used for gantry angle selection (section 2.2), with the same 36 ×36 $cm^2$ fields and zero-degree collimator angle. These measurements were related to corresponding TRS-398 reference dosimetry measurements through the use of ratios.

B-factors were measured for each TSET electron beam, by measuring dose around the surface of a 30 cm diameter phantom (as recommended by AAPM [3]) using radiochromic film, and calculating the ratio of this result to the extended-SSD measurement above.

### 2.4 Patient position corrections

Correction factors to account for displacements of the patient surface position away from the reference distance were measured by starting with the Roos chamber in the TSET reference setup shown in Figure 1, with RW3 White Water buildup included to move the measurement point to the depth of maximum dose, and then moving the phantom containing the chamber by various distances towards and away from the gantry. By taking readings at a range of positions and normalizing the results to the reference distance, a relationship for calculating patient position calculations was determined for each electron beam energy.

### 2.5 Depth-dose measurements

Two different methods were used for depth-dose measurements for the 6 and 9 MeV high-dose-rate electron beams. Firstly, to establish agreement with linacs used previously in the department, measurements of percentage depth ionisation (PDI) were made using anterior beams and beam-pairs only, using the Roos chamber setup described in section 2.1, with the exception that sheets and blocks of Virtual Water (Standard Imaging Inc, Middleton, USA) were used instead of Goettingen RW3 White Water. This change was largely due to known water equivalence of Virtual Water for electron beam measurements. Once the PDI measurements were completed, Burns et al's method [9] was used to convert the results to percentage depth dose (PDD) data.

Secondly, to provide a set of reliable prescription depth correction factors to use in TSET treatment Monitor Unit calculations, and to provide an indication of the dose falloff characteristics from full TSET treatments, an entire 12 beam (i.e. six paired beams) treatment was delivered to a stack of two 30 cm diameter TomoTherapy cheese phantoms (TomoTherapy Inc, now Accuray Inc, Sunnyvale, USA) with radiochromic film placed between, with the film positioned level with isocenter height. The phantom was carefully rotated by 60 degrees between the delivery of each beam pair. This treatment PDD measurement was completed for both the 6 MeV high-dose-rate electron beam and the 9 MeV high-dose rate electron beam.

### 2.6 Vertical and horizontal profile measurements

Vertical (sup-inf) and horizontal (left-right) profile measurements were made using Roos chamber in the same starting setup as is used for the other tests in this paper (described in section 2.1), with sheets of RW3 White Water added to place the chamber's measurement point at the depth of maximum dose.

The phantom block containing the Roos chamber was initially placed at the vertical height and lateral position of the linac isocenter, and then integrated charge readings were obtained after shifts in the vertical and then the horizontal directions. A comprehensive set of data, using 10 cm shifts, was obtained for both the 6 and 9 MeV high-dose rate electron beams.

### 2.7 End-to-end tests

End-to-end testing was performed for both TSET beams using the humanoid phantom shown in Figure 1, an Alderson Rando phantom from 1974 (now supplied as the Alderson Radiation Therapy Phantom, by Radiology Support Devices Inc, Gardena, USA). Surface dose measurements were made using small pieces of radiochromic film taped at reference locations (sternum, umbilicus, upper and lower back) as well as in challenging locations at the side of the hip, where the phantom is widest (closest to gantry) and between the phantom's legs where the radiation line-of-sight is largely obscured. Large pieces of radiochromic film were also cut to size and used to make transverse dose measurements through the phantom's abdomen, to provide a qualitative indication of dose homogeneity and falloff for the 6 and 9 MeV high-dose-rate beams.

The phantom was treated as a patient would be, using six beam-pairs delivered at each of six equidistant phantom rotation angles. For this test, all twenty-four beams (six beam-pairs at each of the two beam energies) were delivered using a fixed 1500 Monitor Units (MU), and the expected surface dose from this MU setting was calculated using the factors described throughout this paper and used as the prescription dose. Film measurement results were evaluated in terms of their level of agreement with this nominal prescription.

## 3 Results

### 3.1 Optimal gantry angles

After some trial and error, the required thicknesses of RW3 White Water needed to achieve a maximum dose measurement, in the 36 ×36 $cm^2$ TSET fields at extended SSD behind the 6 mm spoiler screen were 5 mm for the 6 MeV beam and 12 mm for the 9 MeV beam. Note that RW3 White Water thicknesses were tested in 1 mm increments, and that the water-equivalence of RW3 White Water was not established in this work. (Water-equivalence was not needed so much as a thickness of plastic that would permit a measurement at the maximum dose depth for each beam.)

The iterative gantry angle testing process identified optimal gantry angles 16.9 degrees above and below the horizontal, equating to a hinge angle of 33.8 degrees, for the 6 MeV high-dose-rate electron beam and 16.1 degrees above and below the horizontal, equating to a hinge angle of 32.2 degrees, for the 9 MeV high-dose-rate electron beam.

### 3.2 Reference output factors and B-factors

Reference output factors, accounting for differences between the setup of TRS-398 reference dosimetry measurements [7] and the setup of the TSET extended-SSD reference measurements (section 2.3), were measured as 0.048 for the 6 MeV high-dose-rate electron beam and 0.055 for the 9 MeV high-dose-rate electron beam. These factors being substantially less than 1.000 indicate the extent to which the effects of the increased SSD dominate over the effects of the increased field size, for each beam, compared to conventional reference dosimetry conditions.

B-factors for the 30 cm diameter phantom were measured at 2.70 for the 6 MeV high-dose-rate electron beam and 2.71 for the 9 MeV high-dose-rate electron beam, similar to the B-factor of 2.63 published by Schiapparelli et al [8] and within the typical range 2.5 to 3.1 suggested by the AAPM [3]. The B-factors for the 6 and 9 MeV high-dose-rate electron beams being close to identical to each other highlights the extent to which the relationship between mean surface dose and TSET calibration point dose is determined by treatment geometry rather than the characteristics of the incident radiation beam.

### 3.3 Patient position correction factors

Figure 2 illustrates the results of the patient position correction measurements for the 6 and 9 MeV high-dose-rate electron beams, showing that the two different beams require almost identical correction factors, with linear fits to the measurement data closely overlying each other. Numerically, as the patient position moves away from the reference position, 3000 mm from isocenter, the correction factor deviates from 1.00 by 0.076% per mm and 0.078% per mm for the 6 and 9 MeV high-dose-rate electron beams, respectively.

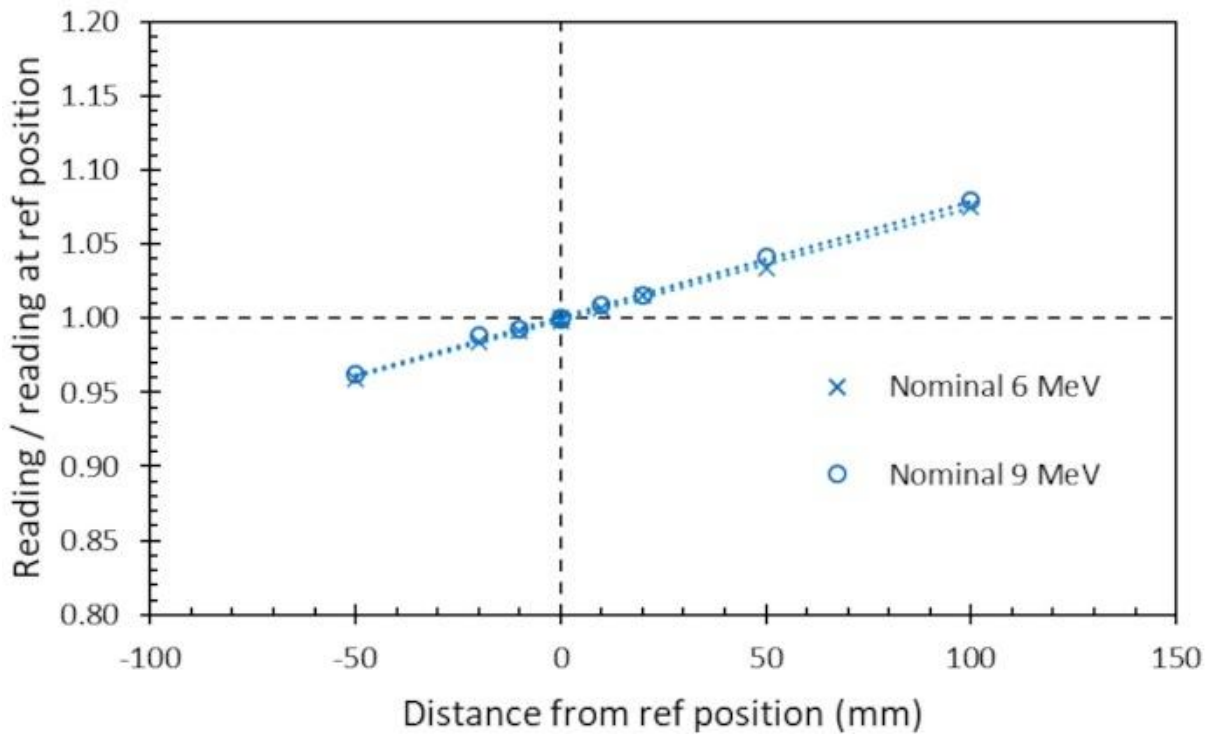

**Fig. 2.** Patient position correction measurement results for 6 and 9 MeV high-dose-rate electron beams, showing linear fits to data. Positive direction is towards gantry.

### 3.4 Depth-dose results

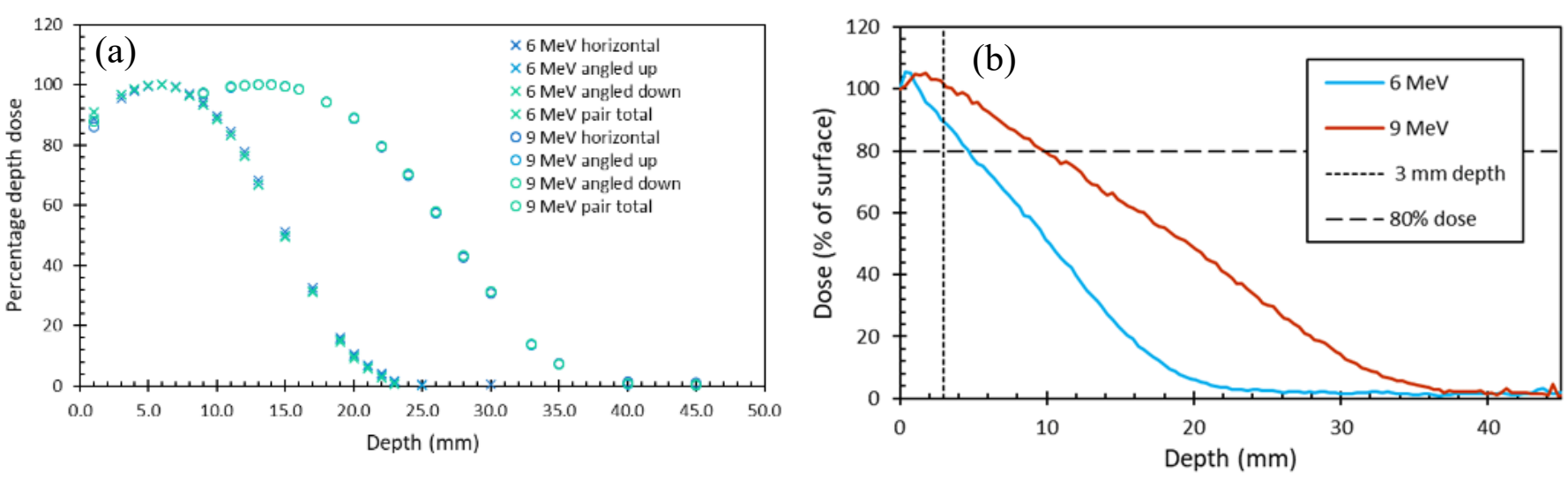

**Fig. 3.** PDD results derived from (a) Roos chamber measurements for anterior beams and beam pairs only, normalized to maximum, and (b) radiochromic film measurements of entire 6-beam-pair treatments normalized to surface, using nominal 6 MeV and 9 MeV electron beams.

Figure 3(a) and 3(b) shows comparisons of 6 MeV and 9 MeV high-dose-rate electron beam measurements of PDDs made in the TSET treatment setup. Figure 3(a) shows results from anterior beams and beam-pairs only, which produce depth-dose curves that are similar in shape to PDDs measured at isocentre. The main difference between these measurement results and the results that would be seen at isocentre is the substantially reduced beam energies and penetration depths, arising from the use of an extended SSD and a 6 mm thick beam spoiler in the TSET setup.

Figure 3(b) shows PDDs measured with radiochromic film, when entire 6-beam-pair treatments are delivered using the 6 MeV and 9 MeV high-dose-rate electron beams. The contributions from dose delivered to the phantom at different orientations sum to give the familiar quasi-linear falloff that is a familiar feature of TSET treatments [3, 8]. On this plot, the dose levels at 3 mm depth (a common TSET prescription depth) and the depths of the 80 % isodoses (recommended to fall between 7 and 10 mm according to lymphoma consensus guidelines [10]) are highlighted for comparison.

### 3.5 Vertical and horizontal profiles

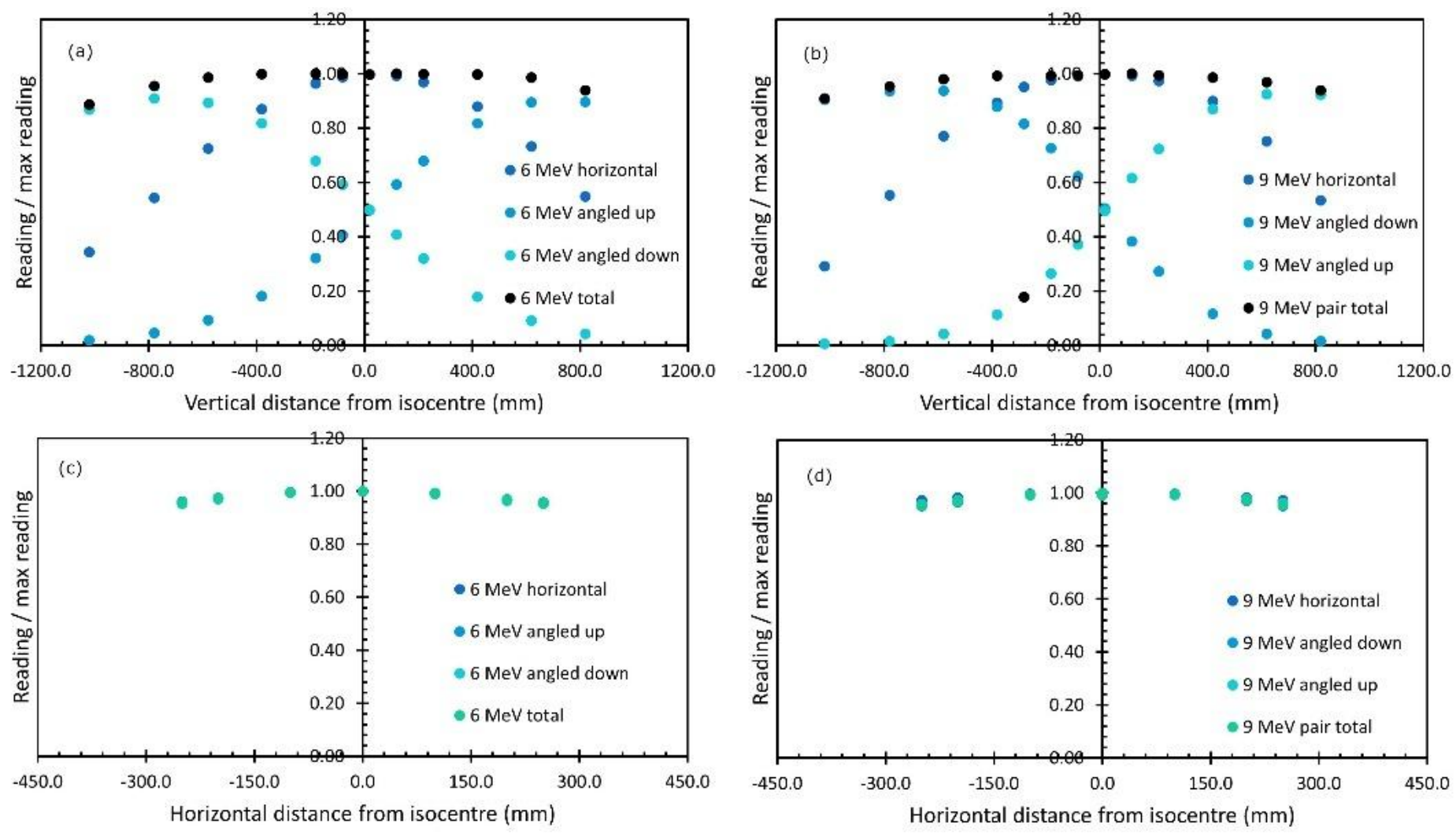

**Fig. 4.** (a) and (b): vertical profiles, (c) and (d): horizontal profiles measured in TSET setup, for nominal 6 MeV beam ((a) and (c)) and nominal 9 MeV beam ((b) and (d)).

Figures 4(a), (b), (c) and (d) show the vertical (superior-inferior) and horizontal (left-right) profile measurement results for the 6 and 9 MeV high-dose-rate electron beams, delivered to a Roos chamber measurement assembly that was iteratively shifted around the patient plane in the TSET treatment setup. All four results show the effects of the 6 mm spoiler screen, in achieving dose homogeneity throughout the measurement region. Figures 4(a) and (b) also highlight the advantage of using two paired beams, one angled upward and the other angled downward, for achieving a vertical dose homogeneity substantially greater than what is evidently achievable using a horizontal beam alone.

Comparisons between these results show that equivalent homogeneity is achieved using the 6 and 9 MeV high-dose-rate electron beams, with both beams achieving a dose homogeneity within 2% over a total horizontal distance of 600 mm and a total vertical distance of 1400 mm.

### 3.6 End-to-end testing results

Figure 5 provides an indication of the setup and results of the end-to-end tests that were performed for the 6 and 9 MeV high-dose-rate electron beams. Visual examination of the transverse films indicated that the expected level of dose homogeneity was achieved, around the surface of the phantom, for both treatments. There was also an obvious dose falloff resulting in minimal film darkening throughout the interior of the phantom, although the falloff for the 9 MeV beam was substantially reduced compared to the 6 MeV beam (as expected given the results shown in Figure 3).

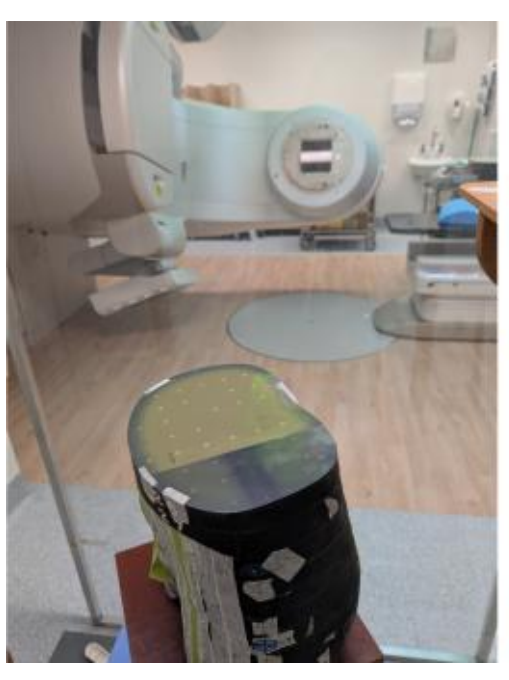

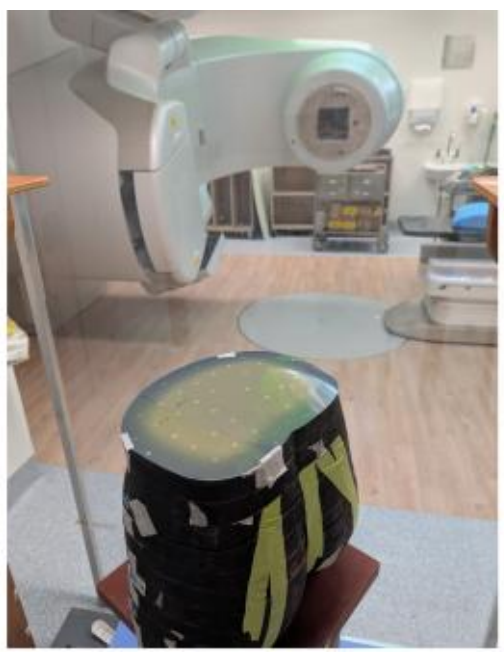

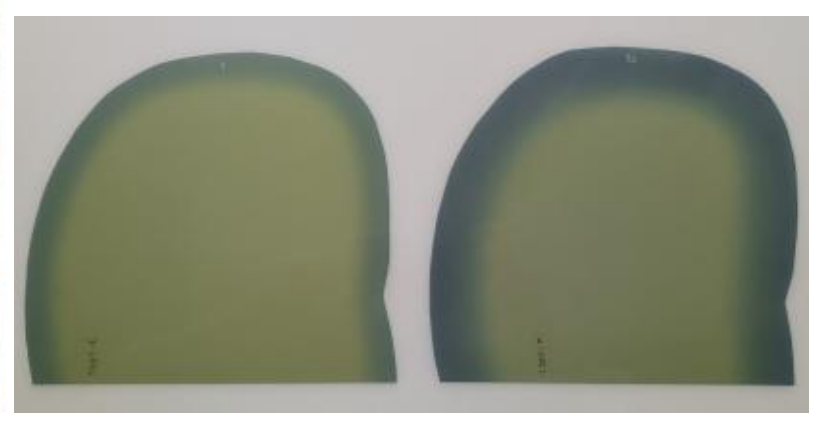

**Fig. 5.** Photographs taken at the completion of TSET end-to-end testing, showing transverse films in situ and after removal from the phantom. Note that scrap film was used for part of the 6 MeV measurement, as this was largely a spot-check. Note also that the rest of the humanoid phantom was placed on top of the film, during both treatment deliveries.

The mean dose measurements at the reference points on the phantom (sternum, umbilicus, upper and lower back) agreed with the prescribed surface dose within 2% for both beam energies, as did the dose measured at the widest part of the phantom's hip. As expected, dose between the phantom's legs was substantially reduced, compared to the prescription, measuring colder than the prescription by 61% for the 6 MeV high-dose-rate beam, echoing the results of in vivo measurements for patients treated with this beam [6]. For the evidently more penetrating 9 MeV beam, the dose in this challenging region was decreased by only 46% compared to the prescribed surface dose.

## 4 Discussion

Table 1 lists the key features of the 6 and 9 MeV high-dose-rate electron beams, evaluated under TSET conditions, for the purpose of evaluating the utility of the latter beam as a modality suitable for treating unusually deep and extensive mycosis fungoides lesions.

The key clinical distinction between the two beams is evidently the increased depth penetration from the 9 MeV treatment beam pair. The 80% dose depth increased to nearly 10 mm and the practical range increased to 35 mm, while the maximum dose remained within 2 mm of the surface (see Figure 3(b)). Although the 9 MeV beam was subject to less air-scatter (leading to a reduced hinge angle), vertical and horizontal dose profiles measured with the two beam-pairs were almost indistinguishable (see Figure 4(a)-(d)). Dose from both beam-pairs was homogeneous (within 2% of central axis result) over a total vertical distance of 140 cm.

These two beams are referred to as "6 MeV" and "9 MeV" throughout this paper, for simplicity only. Calculating the mean surface dose for these beams when they reach the phantom at extended SSD behind a 6 mm spoiler screen, using the PDD data shown in Figure 3(a) via a method proposed specifically for evaluating TSET beams [3], results in $E_0$values of 3.4 MeV for the nominal 6 MeV beam, and 6.3 MeV for the nominal 9 MeV beam.

The move to a higher energy TSET treatment beam, for the rare cases where it may be needed, may require a re-evaluation of any shielding materials that are used for less penetrating TSET beams, e.g., for eye, finger/toenail and scrotal shielding [6] and occasionally for cranial (i.e. hair follicle) shielding. Ideally the dose calculation method, for a beam that penetrates up to 35 mm into the patient's tissue, would also be more sophisticated than the manual calculation techniques that are commonly used today [5].

**Table 1.** 6 and 9 MeV high-dose-rate electron beam characteristics measured at extended SDD behind spoiler screen (significant figures indicate uncertainties).

| Result | Nominal 6 MeV electron beam | Nominal 9 MeV electron beam |
|---|---|---|
| Extended-SSD large field reference dose / isocenter reference dose (ratio) | 0.048 | 0.055 |
| Depth of max dose for horizontal beam (mm) | 6 | 13 |
| $R_{50}$ for horizontal beam (mm) | 15 | 27 |
| Mean energy at phantom surface ($E_0$) for horizontal beam (MeV) | 3.4 | 6.3 |
| Optimal gantry angles (° from horizontal) | +/- 16.9 | +/-16.1 |
| Optimal hinge angle (°) | 33.8 | 32.2 |
| Depth of max dose for paired treatment beams (mm) | 0.5 | 1.8 |
| Depth of 80% dose for paired treatment beams (mm) | 4.7 | 9.7 |
| Max dose from paired treatment beams (% of surface dose) | 105.5 | 105.0 |
| Dose at 3 mm depth from paired treatment beams (% of surface dose) | 89.8 | 102.2 |
| Practical range of paired treatment beams (mm) | 19 | 35 |
| Bresmstrahlung tail from paired treatment beams (% of surface dose) | 1.35 | 1.37 |
| Homogeneous dose region from paired treatment beams, vertical (mm) | +/- 700 | +/- 700 |
| Homogeneous dose region from paired treatment beams, horizontal (mm) | +/- 300 | +/- 300 |
| Patient position correction factor rate of deviation from 1.00 (% per mm) | 0.076 | 0.078 |
| B factor for 30 cm diameter phantom (ratio) | 2.70 | 2.71 |
| End-to-end mean reference measurement diff from prescribed surface dose (%) | 1.8 | -0.5 |
| End-to-end worst case (between legs) measurement diff from prescribed surface dose (%) | -61 | -46 |

# 5 Conclusion

The 9 MeV TSET beam has been shown to achieve greater depth penetration compared to the commonly used 6 MeV TSET beam, without detrimentally affecting the dose uniformity achievable in the patient plane. TSET treatments with 9 MeV electrons may be advisable for patients with extensive lesions that are too deep for effective treatment with a 6 MeV beam, although a re-evaluation of shielding materials and dose calculation methods would also be advisable.

# References

1. Piotrowski, T., Milecki, P., Skórska, M., Fundowicz, D.: Total skin electron irradiation techniques: a review. Postepy Dermatol Alergol. 30, 50-55 (2013). DOI:10.5114/pdia.2013.33379
2. Diamantopoulos, S., Platoni, K., Dilvoi, M., Nazos, I., Geropantas, K., Maravelis, G., Tolia, M., Beli, I., Efstathopoulos, E., Pantelakos, P. and Panayiotakis, G.: Clinical implementation of total skin electron beam (TSEB) therapy: a review of the relevant literature. Phys Med. 27, 62-68 (2011). DOI:10.1016/j.ejmp.2010.09.001
3. Karzmark, C.J., Anderson, J., Buffa, A., Fessenden, P., Khan, Svensson, G., Wright, K.: AAPM Report No. 23 - Total skin electron therapy: technique and dosimetry. American Association of Physicists in Medicine (1987). DOI:10.37206/22
4. Fitchew, R., Nitschke, K., Christiansen, P.: Total skin electron beam therapy using a Dynaray 18 linear accelerator. Australas Phys Eng Sci Med. 8, 182-187 (1985).
5. Crowe, S., Cassim, N., Livingstone, A., Luscombe, J., McMahon, K., Kairn, T.: Total skin electron therapy dose calculations using photogrammetry-derived synthetic CT data. Phys Eng Sci Med. 47, 1854 (2024). DOI:10.1007/s13246-024-01460-7
6. Kairn, T., Wilks, R., Yu, L., Lancaster, C., Crowe, S.B.: In vivo monitoring of total skin electron dose using optically stimulated luminescence dosimeters. Rep Pract Oncol Radiother. 25, 35-40 (2020). DOI:10.1016/j.rpor.2019.12.011
7. IAEA Technical Reports Series No. 398 (Rev. 1) - Absorbed dose determination in external beam radiotherapy: An international code of practice for dosimetry based on standards of absorbed dose to water. International Atomic Energy Agency (2024). DOI:10.61092/iaea.ve7q-y94k
8. Schiapparelli, P., Zefiro, D., Massone, F., Taccini, G.: Total skin electron therapy (TSET): a reimplementation using radiochromic films and IAEA TRS-398 code of practice. Med Phys. 37, 3510-3517 (2010). DOI:10.1118/1.3442301
9. Burns, D.T., Ding, G.X., Rogers, D.W.: R50 as a beam quality specifier for selecting stopping-power ratios and reference depths for electron dosimetry. Med Phys. 23, 383-388 (1996). DOI:10.1118/1.597893
10. Specht, L., Dabaja, B., Illidge, T., Wilson, L.D., Hoppe, R.T.: International Lymphoma Radiation Oncology Group. Modern radiation therapy for primary cutaneous lymphomas: field and dose guidelines from the International Lymphoma Radiation Oncology Group. Int J Radiat Oncol Biol Phys. 92, 32-39 (2015). DOI:10.1016/j.ijrobp.2015.01.008